\documentclass[lettersize,journal]{IEEEtran}
\usepackage{amsmath,amsfonts}
\usepackage{algorithmic}
\usepackage{algorithm}
\usepackage{array}
\usepackage[caption=false,font=normalsize,labelfont=sf,textfont=sf]{subfig}
\usepackage{textcomp}
\usepackage{stfloats}
\usepackage{url}
\usepackage{verbatim}
\usepackage{graphicx}
\usepackage{cite}
\usepackage{multirow}
\usepackage{multicol}
\usepackage{booktabs}
\usepackage[algo2e]{algorithm2e} 
\usepackage[inline]{enumitem}
\usepackage{csquotes}

\begin{document}

\title{Human and social capital strategies for Mafia network disruption}

\author{Annamaria Ficara, Francesco Curreri, Giacomo Fiumara, Pasquale De Meo}



\maketitle

\begin{abstract}
Social Network Analysis (SNA) is  an  interdisciplinary  science  that  focuses  on  discovering  the  patterns  of  individuals interactions. In particular, practitioners have used SNA to describe and analyze criminal networks to highlight subgroups, key actors, strengths and weaknesses in order to generate disruption interventions and crime prevention systems.
In this paper, the effectiveness of a total of seven disruption strategies for two real Mafia networks is investigated adopting SNA tools. Three interventions targeting actors with a high level of social capital and three interventions targeting those with a high human capital are put to the test and compared between each other and with random node removal. Human and social capital approaches were also applied on the Barab\'{a}si-Albert models which are the one which better represent criminal networks. Simulations showed that actor removal based on social capital proved to be the most effective strategy, by leading to the total disruption of the criminal network in the least number of steps. The removal of a specific figure of a Mafia family such as the Caporegime seemed also promising in the network disruption.\end{abstract}

\begin{IEEEkeywords}
Criminal Network, Social Network Analysis, Disruption, Social Capital, Human Capital, Simulation.
\end{IEEEkeywords}

\section{Introduction}
\label{intro}

\IEEEPARstart{C}{riminal} organizations are groups that covertly engage in illegal activities to provide goods and services to gain a profit, by accomplishing achievements at the cost of other individuals, groups or societies~\cite{Finckenauer2005}. In particular, Mafia is a criminal group defined by Gambetta as a ``territorially based criminal organization that attempts to govern territories and markets"~\cite{gambetta1996sicilian}, by defining the one located in Sicily as the \textit{original Mafia}. Compared to other criminal organizations, Mafia groups differ in their structure. They are structured as a collection of loosely coupled groups, which last for several generations~\cite{Mastrobuoni2012, sciarrone2014territorial}. Each of these groups is referred to as \textit{cosca}, \textit{family} or \textit{clan}.
Because of their strong resilience to disruption, such networks pose particularly hard challenges to Law Enforcement Agencies (LEAs). Herein, we borrow methods and tools from Social Network Analysis (SNA) to investigate the effectiveness of several law enforcement interventions against two Mafia networks, based on a real-world dataset built from a major anti-mafia operation called ``Montagna" which was concluded in 2007. Such dataset was used in different studies on Mafia networks through SNA, in particular to analyze the structure of such networks~\cite{Ficara2020, Cavallaro2021,Ficara2021Green}, identifying subgroups and highlighting strategically positioned key actors~\cite{ficara2021multilayer, FicaraCompleNet2021}, and developing disruption and prevention methods~\cite{ficara2021criminal,cavallaro2020disrupting, Calderoni2020}.

SNA is a growing interdisciplinary science that focuses on discovering the patterns of individuals interactions.  It found extensive application in organizational behavior, inter-organizational relations, criminal groups analysis, the study of breakouts, mental health, the diffusion of information. As an interdisciplinary science, it spans through different domains such as Anthropology, Sociology, Psychology, Economics, Mathematics, Medicine and Computer Science~\cite{camacho2021new}. Some of the challenges currently involving SNA deal with big data analytics, information fusion, scalability, statistical modeling for large networks, pattern modeling and extraction, or visualization.


The idea of conceiving organized crime as a network, rather than a hierarchical structure, has incrementally grown in criminologist literature over the last century. During the twentieth century, the most common approach to study organized crime was the ``alien conspiracy theory", that blamed the origin of crime to outsiders (hence its name) and that considered it structured as a bureaucratic organization that followed a specific hierarchy with specific roles~\cite{CRJ455}. Only by the end of that century, such view was abandoned in favor of new analytical methods that viewed organized crime as a system of loosely structured relationships mainly based on patron-client relations~\cite{albini1971american}. Investigations started to be conducted by performing link analysis through visual representations of the structure of the criminal groups~\cite{Lupsha1983NetworksVN}, giving birth to the first applications of network analysis as a ``tool"~\cite{ianni1973ethnic}. Already by the '80s, network methods and concepts, such as density and centrality, were adopted to study criminal groups~\cite{Lupsha1983NetworksVN} and time by time the interest in the discipline grew significantly, contributing to opening new research trends and bringing significant developments~\cite{ianni1990network, SPARROW1991251}.

Nowadays, SNA for criminal analysis focuses on the computation for network measurements such as density and centralization, analysis of clusters and the measure of the centrality of individuals, that are able to identify critical actors in the group~\cite{Alzaabi2015,Taha2016,Taha2017,Taha2019} and whose removal would maximize network disruption~\cite{MorselliPetit2007,Duijn2014} and allow to construct crime prevention systems~\cite{Calderoni2020,berlusconi2016link}.
Some actors within criminal networks can have particularly strong connections or more connections than others. Some actors may also act as brokers between other actors. 

Social networks, including criminal networks, can thus be conceptualized as being made of two kinds of \textit{capital}, on which dismantling approaches are based: human capital and social capital. The first approach deals with techniques able to identify high human capital figures within the network and it is based on the importance of roles. Human capital is indeed defined as ``the knowledge, skills, competencies and attributes embodied in individuals that facilitate the creation of personal, social and economic well-being"~\cite{Keeley2007}. The social capital approach identifies key actors based on their ties in the group and the centrality measures from SNA, giving more importance to the communicative flow rather than the importance of roles. Social capital is defined as ``those tangible assets that count for most in the daily lives of people: namely goodwill, fellowship, sympathy, and social intercourse among the individuals and families who make up a social unit"~\cite{Keeley2007}.

In~\cite{math10162929}, we provided a literature review of the most significant works on covert network disruption underlining how disruption strategies based on human or social capitals are usually developed in a parallel fashion and exploit very different techniques. Social and human capital can be used together thus creating a third approach which we defined as the mixed approach. Only few researchers tried this unified approach that seeks to identify nodes in the network which are simultaneously able to deteriorate both of the capitals~\cite{Carley2001, Krebs02mappingnetworks, Tsvetovat2005, morselliglance, Robins2009, Spapens2010, Nguyen2013, Duijn2014,Duijn2014book,  Bright2017,Villani2019}. For example, Bright \textit{et al.}~\cite{Bright2017} adopted different strategies based on either human or social capital exploring the validity of five LEAs interventions in dismantling and disrupting criminal networks.
We also noticed that most of the works on covert network disruption refer to terrorist networks while there are very few papers about Mafia network disruption. 
For this reason and inspired by~\cite{Bright2017}, in this paper, we introduced and tested on our Mafia networks different disruption strategies based on either human and social capital. We used three traditional centrality measures to target actors with a high level of social capital: degree, betweenness and closeness centralities. We also selected caporegimes, soldiers and entrepreneurs as targeted actors with a high level of human capital. At each step, an actor is removed and the network integrity is evaluated through three measures, namely the number of connected components, the size of the largest connected component and the average global efficiency. Such interventions are then compared with each other and with random removal of actors for both of the two networks, leading to a total of seven different strategies. The random strategy is an important comparison since it can be seen as the case in which LEAs randomly raid sites of illegal actions making arrests on the spot.
The aim is to gain insights about the most efficient disruptive strategy, that is evaluated by the number of steps before the complete disruption and consistency above replications.

Bright \textit{et al.}~\cite{Bright2017} considered a case study related to the manufacture and trafficking of synthetic drugs like methamphetamine. To perform their duties and carry out the process of drug cultivation, production, and distribution, the actors within the network must possess specific resources: drugs, precursor chemicals, equipment, money, premises, skills, labor and information. The used human capital strategy was implemented by removing actors possessing a particular resource and with the highest degree centrality value. Our approach is different because our aim is to dismantle mafia families, which have a very peculiar hierarchical structure~\cite{ficara2021multilayer}, with different characteristics from simple criminal groups who produce and distribute synthetic drugs. In fact, our human capital strategy does not consist in targeting actors who possess a particular resource but those who have a particular role in the hierarchy of the mafia group.
Then, we want to verify if it is possible to create a network model for criminal network disruption using an artificial network with the same characteristics of a mafia network.
In our previous work~\cite{Ficara2021Green}, we used some popular network models like the Erd\"{o}s-R\'{e}nyi (ER)~\cite{Erdos1959} model, the Watts-Strogatz (WS)~\cite{Watts1998} model and different configurations of the Barab\'{a}si-Albert (BA)~\cite{Barabasi1999} model to replicate the topology of a criminal network. Our experiments identified the BA model as the one which better represents a criminal network. Once we have identified the key role in the hierarchy of a mafia family or in its criminal activities, we want to try to identify this role in BA models and apply our disruption strategies to these models. Specifically, the human capital approach is simulated targeting nodes with the same rank of the caporegimes in our Mafia networks.

The paper is structured as follows. In Section~\ref{sec:1}, 
our dataset adopted in this work is introduced; Section~\ref{sec:distrat} describes all the seven strategies adopted, divided in three subgroups: social capital based strategies, human capital based strategies and random disruption; the algorithms for the simulations are explained and measures to evaluate the network integrity are given as well; Section~\ref{sec:3} shows the results and comparison between the methods applied on Mafia networks and BA models and conclusions are finally drawn.


\section{Criminal network dataset}
\label{sec:1}

Our analysis focuses on two real criminal networks related to a specific anti-mafia operation called Montagna~\cite{Calderoni2020, cavallaro2020disrupting, ficara2021criminal, Ficara2021Green, ficara2021multilayer,FicaraCompleNet2021}. 

This operation was conducted by the Special Operations Group (ROS) of the Italian Carabinieri and the Provincial Command of Messina (Sicily) who were able to eliminate leaders of the Mistretta Mafia family and the Batanesi clan (operating in Tortorici) making 39 arrests under preventive detention orders
and reporting 28 suspected criminals on the loose. The Mistretta family and the Batanesi clan, between 2003 and 2007, monopolized the sector of public contracts in the Tyrrhenian strip and in the nebroidal district of the province of Messina, through a cartel of entrepreneurs close to the Sicilian Mafia. Between the end of the '90s and the beginning of the 2000s, these entrepreneurs acquired important public orders, from supplies for works on roads and highways to contracts for the methanization of many municipalities in the area. Furthermore, the Montagna operation identified the Mistretta family as a mediator between Mafia families in Palermo and Catania and other criminal organizations around Messina. 

\begin{figure*}[!t]
\centering
\includegraphics[width=\textwidth]{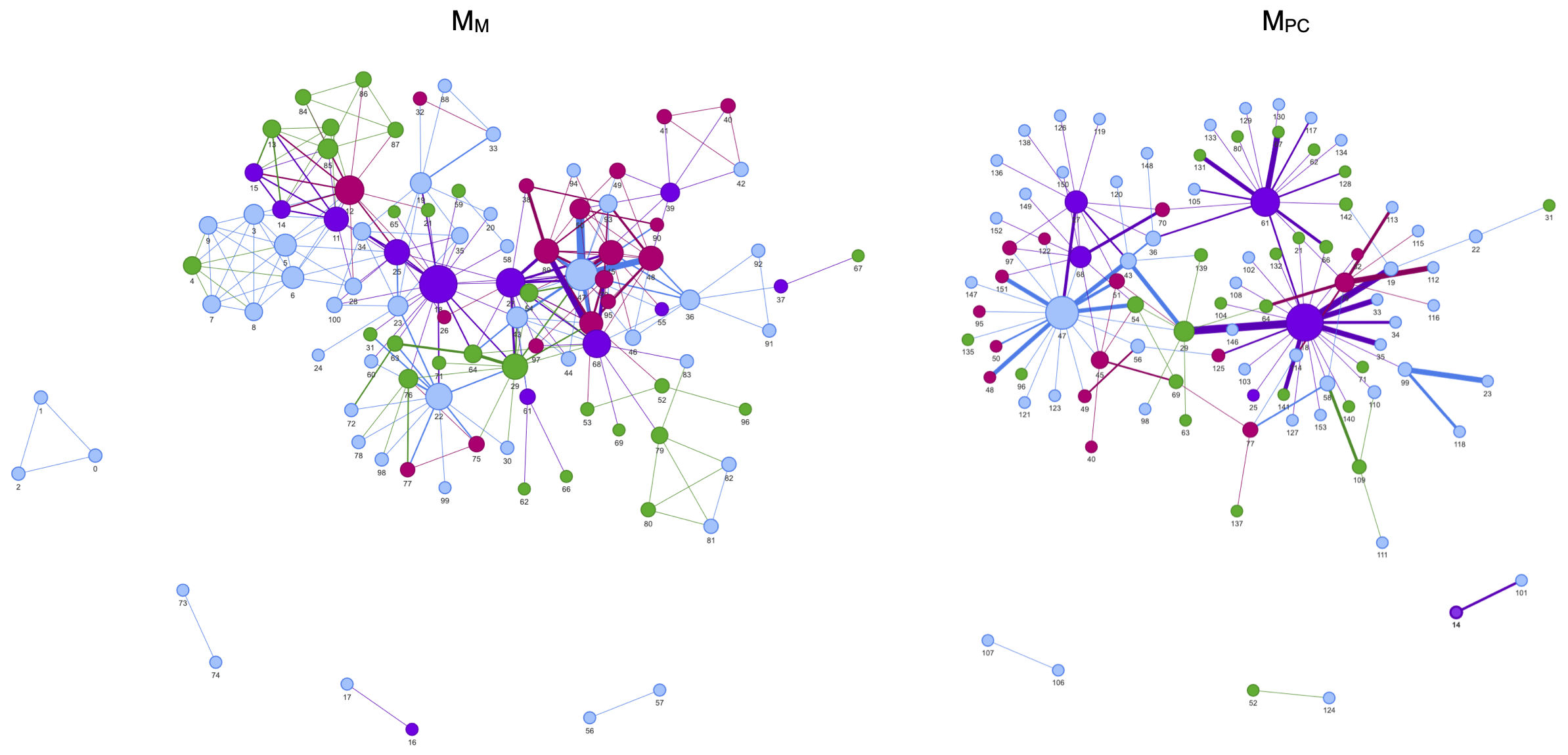}
\caption{The Montagna Meetings $M_M$ and Phone Calls $M_{PC}$ networks. The edge width is equal to the edge weights, \textit{i.e.} the number of meetings (or phone calls). The node size is proportional to node degree. Caporegimes are marked in purple, soldiers in burgundy and entrepreneurs in green.}
\label{fig:nets}
\end{figure*}

In 2007, after the conclusion of the anti-mafia operation, a pre-trial detention order for 38 individuals was issued by the Preliminary Investigation Judge of Messina. It was a two hundred pages document which contained a lot of details about crimes, activities, meetings, and calls among the suspected criminals. From this order, we extracted two unique undirected and weighted networks, \textit{i.e.} Montagna Meetings $M_{M}$ and Montagna Phone Calls $M_{PC}$. The first one contains $101$ suspected criminals close to the Sicilian Mafia connected by $256$ links which represent meetings emerging from the police physical surveillance. The second one contains $100$ suspects connected by $124$ links which represent phone calls emerging from the police audio surveillance. $M_{M}$ and $M_{PC}$ share $47$ nodes and are available on Zenodo~\cite{Zenodo2020}.


As we have already discussed in~\cite{ficara2021multilayer}, a Mafia family or clan has a typical hierarchical structure. On top of the pyramid hierarchical chart is the Boss who keeps a low-profile often hiding his real identity. He makes all the major decisions, controls the other members of the clan and resolves any kind of dispute. Just below him is the Underboss who is the second in command. If the Boss risks going to jail or is pretty old, the Underboss can replace him and resolve some disputes without involving him.
In-between the Boss and Underboss there are two key roles which are the Consigliere and the Messaggero. The first one advices the boss and makes fair decisions for the good of the Mafia. The second one is a messenger who limits the public exposure of the boss, reducing the need for sit-downs or meetings between the clans.
In a specific geographical location, the Caporegime or Capo manages his group of criminals within the family. He is just below the underboss and his career depends on the amount of money he can bring into the criminal family. The number of Caporegimes in a given family depends on the dimension of that family. A capo can have many soldiers in his crew. Soldiers are street level mobsters who essentially are no more than average criminals. Then come associates who work with Mafia soldiers and caporegimes on various criminal activities. They can be drug dealers or thieves, as well as entrepreneurs, pharmacists, lawyers, politicians or police officers, who are not actual members of the Mafia, but work with the mob. 

\begin{table}[!t]
\centering
\caption{List of attributes and number of nodes who possess each in the Meetings and Phone Calls networks extracted from the Montagna operation\label{tab1}}
\begin{tabular}{|ll|cc|}
\toprule
\multirow{2}{*}{Attribute} & & \multicolumn{2}{c|}{No. nodes}\\
\cmidrule{3-4}
 & & Meetings & Phone Calls\\
\midrule
Boss & & 4 & 0 \\
Messaggero & & 1 & 1\\
Caporegime & & 12 & 7 \\
Deputy Caporegime & & 2 & 2 \\ 
Soldier & & 18 & 18 \\
\multirow{13}{*}{Associate} & Entrepreneur & 26 & 25 \\ 
& Pharmacist & 2 & 2 \\ 
& Lawyer & 1 & 1 \\
& Electrician & 1 & 0 \\
& City employee & 0 & 1 \\
& Transporter & 0 & 2 \\
& Cooperating witness & 1 & 0 \\
& Landowner & 0 & 1 \\
& Bar owner & 0 & 1 \\
& Fishmonger & 0 & 1 \\
& Accountant & 0 & 1 \\
& Breeder & 2 & 1 \\ 
& Construction worker & 1 & 0 \\
& External partnership & 5 & 8 \\
Relative & & 6 & 3 \\
Cohabitee & & 0 & 2 \\
Fugitive & & 1 & 0 \\
Charged & & 0 & 2 \\
In jail & & 2 & 3 \\
Figurehead & & 0 & 2 \\
Unclear & & 16 & 16 \\
\bottomrule
\end{tabular}
\end{table}

Starting from our pre-trial detention order, we were also able to reconstruct the roles of the actors according to the specific hierarchy of Mafia families and also defining the roles of associates in our criminal networks. Thus, we built a labeled graph in which each node has an attribute as described in Table~\ref{tab1}.

Crimes committed by the Mafia families at the centre of the Montagna operation involve flow of human capital, resources, information, specific roles and tasks.
The identification of specific roles is important for the development of human capital strategies for network disruption. For example, the role of an associate as an entrepreneur could be important to win public tenders and to accomplish the public contracts in a fraudulent way. Also soldiers can be important because they are those who actually commit crimes such as beatings, money collection and robbery. Then, caporegimes have a significant role having the major social status and influence in the organization. They command a crew of soldiers and report directly to the Boss or the Underboss. In Figure~\ref{fig:nets}, caporegimes, soldiers and entrepreneurs are colored in purple, burgundy and green, respectively.


\section{Criminal network disruption strategies}
\label{sec:distrat}

In our experiments we reproduced the interventions that law enforcement agencies usually carry out to disrupt and dismantle criminal networks, that is to say we removed a node and all the incident edges. The nodes were selected according to their human and social capital, following criteria that will be discussed in detail in Subsects.~\ref{social} and~\ref{human}.
Each of these interventions was modeled by a targeting method which begins with the full networks $M_M$ and $M_{PC}$ respectively of $101$ and $100$ actors. At each time step we delete a node according to the specific targeting method.  At each step we measured the number of connected components, the size of the largest connected component, and the average global efficiency. For each intervention, the simulation stopped when the network was completely disrupted: that is, when no nodes remain. 
For this study three general disruption approaches have been used: social capital disruption, random disruption and human capital disruption and a total of seven different disruption strategies.

\subsection{Social capital disruption} 
\label{social}

The social capital disruption approach aims at strategic positions within criminal networks. It is described in the Algorithm~\ref{alg1}. 

We used three main strategies: degree centrality attack, betweenness centrality attack and closeness centrality attack.

Degree centrality attacks were implemented by removing the actors sequentially according to the maximal degree centrality. Degree centrality (DC)~\cite{FREEMAN1978215} determines the importance of an actor based on the number of connections and it is defined as 
\begin{equation}
DC_i = \frac{d_i}{n-1},
\end{equation}
where $d_i$ is the degree of the actor $i$ and $n$ is the number of nodes of the network. 
High degree centrality actors are called hubs because they are important for the flow of resources and information throughout the network~\cite{Duijn2014}. Hubs are associated with powerful and influential positions within social networks.

Betweenness centrality attacks were implemented by removing the actors sequentially according to the maximal betweenness centrality. Betweenness centrality (BC)~\cite{Brandes} measures how frequently a node lies on the shortest paths between other pairs of nodes: 
\begin{equation}
BC_i = \sum\limits_{h,k} \frac{v^i_{hk}}{g_{hk}},
\end{equation}
where $v^i_{hk}$ is the number of shortest paths from the actor $h$ to the actor $k$ by passing through $i$ and $g_{hk}$ is the total number of shortest paths from $h$ to $k$.
BC represents the ability of some actors to control the flow of connectivity (\textit{e.g.} information, resources etc.) within the network. Since these actors often connect otherwise poorly connected parts of the network, they are called brokers. 

Closeness centrality attacks were implemented by removing the actors sequentially according to the maximal closeness centrality. Closeness centrality (CL)~\cite{FREEMAN1978215} is defined as:
\begin{equation}
CL_i = \frac{n}{\sum\limits_j d_{ij}},    
\end{equation}
where $d_{ij}$ is the distance between $i$ and $j$ and $n$ is the size of the network. 
CL measures how close an actor is to the other actors in the network. This measure has the aim of measuring the ability of autonomy or independence of the actors. 

\begin{algorithm}[ht!]
\SetAlgoLined
{\it \% Initialization;}

set an undirected graph $G = (V, E)$;

set the initial number of connected components $cc_0$ of $G$;

set the initial size of the largest connected component $lcc_0$ of $G$;

set the initial average global efficiency $E^0_{glob}$ of $G$;

set $T = |V|$, the number of steps to stop the algorithm;

\For{each step $s = 1:T$}{

    {\it \% Choose a centrality measure (Degree, Betweenness, Closeness);}
    
    compute the centrality of each node $n \in V$;
    
    {\it \% Apply the target strategy to disrupt $G$;}
   
    set a node $c \in V$ as the most central;
    
    remove $c$ from $V$;

    {\it \% Compute the normalized number of connected components;}
    
    $cc_s = cc_s/cc_0$;
    
    {\it \% Compute the normalized size of the largest connected component;}
    
    $lcc_s = lcc_s/lcc_0$; 
    
     {\it \% Compute the normalized average global efficiency;}
    
    $E^s_{glob} = E^s_{glob}/E^0_{glob}$;
    
}
\caption{Social capital disruption.}
\label{alg1}
\end{algorithm}

\subsection{Random disruption} 
The random disruption approach follows no preference or ranking during the actor selection for removal. It is described in the Algorithm~\ref{alg2}. This strategy can be associated with non-strategic opportunistic law enforcement interventions. This is the case in which for example law enforcement officers randomly bust sites of illicit activities and make arrests on the spot~\cite{Duijn2014}.

\begin{algorithm}[ht!]
\SetAlgoLined
{\it \% Initialization;}

set an undirected graph $G = (V, E)$;

set the initial number of connected components $cc_0$ of $G$;

set the initial size of the largest connected component $lcc_0$ of $G$;

set the initial average global efficiency $E^0_{glob}$ of $G$;

set $T = |V|$, the number of steps to stop the algorithm;

\For{each step $s = 1:T$}{

    {\it \% Apply the random selection strategy to disrupt $G$;}
    
    randomly pick a node $n \in V$;
    
    remove $n$ from $V$;
    
     {\it \% Compute the normalized number of connected components;}
    
    $cc_s = cc_s/cc_0$;
    
    {\it \% Compute the normalized size of the largest connected component;}
    
    $lcc_s = lcc_s/lcc_0$; 
    
     {\it \% Compute the normalized average global efficiency;}
    
    $E^s_{glob} = E^s_{glob}/E^0_{glob}$;

}
\caption{Random disruption.}
\label{alg2}
\end{algorithm}

\subsection{Human capital disruption} 
\label{human}

The human capital disruption strategy consists in targeting actors with special skills or knowledge. This approach is described in Algorithm~\ref{alg3}. 

Based on observations within the data under study and the literature on Mafia networks,
the roles of entrepreneur, soldier and caporegime were selected to analyze this strategy. 

Targeting entrepreneurs attacks were implemented by removing the actors with the specific role of entrepreneur in order of decreasing DC.

Targeting soldiers attacks were implemented by removing the actors with the specific role of soldier in order of decreasing DC.

Targeting caporegimes attacks were implemented by removing the actors with the specific role of caporegime in order of decreasing DC.

\begin{algorithm}[ht!]
\SetAlgoLined
{\it \% Initialization;}

set an undirected graph $G = (V, E)$;

add customize labels on $G$ nodes according to Table \ref{tab1};

 
set $S \subset V$ as a subset of nodes with a specific label {\it (Entrepreneur, Soldier, Caporegime)};

set the initial number of connected components $cc_0$ of $G$;

set the initial size of the largest connected component $lcc_0$ of $G$;

set the initial average global efficiency $E^0_{glob}$ of $G$;

set $T = |S|$, the number of steps to stop the algorithm;

\For{each step $s = 1:T$}{
    
    {\it \% Compute centrality;}
    
    compute the degree centrality of each node $n \in S$;
    
    {\it \% Apply the target strategy to disrupt $G$;}
  
    set a node $c \in S$ as the most central;
    
    remove $c$ from $S$;
    
    {\it \% Compute the normalized number of connected components;}
    
    $cc_s = cc_s/cc_0$;
    
    {\it \% Compute the normalized size of the largest connected component;}
    
    $lcc_s = lcc_s/lcc_0$; 
    
     {\it \% Compute the normalized average global efficiency;}
    
    $E^s_{glob} = E^s_{glob}/E^0_{glob}$;
}
\caption{Human capital disruption.}
\label{alg3}
\end{algorithm}

\subsection{Disruption effects on criminal network structure}

As portrayed in Algorithms~\ref{alg1},~\ref{alg2},~\ref{alg3}, after each actor removal, performed following the disruption strategies described above, we wanted to measure the impact of our attacks on the networks structure in terms of connectivity and efficiency. Therefore we used the following metrics:
\begin{enumerate*}[label={(\arabic*)}]
    \item the number of connected components $cc$;
    \item the largest connected component $lcc$;
    \item the global efficiency $E_{glob}$.
\end{enumerate*}

The connected components show the reachability within the network. In connected components, all the nodes are in fact always reachable from each other. When the number of connected components increases, the number of isolated nodes increases.

In real undirected graphs, we typically find that there is a largest connected component which fills most of the graph while the rest of the network is divided into a large number of small components disconnected from the rest.

Latora and Marchiori~\cite{Latora2001} introduced the concept of efficiency of a graph as a measure of how efficiently it exchanges information. The average efficiency of a pair of nodes $i$ and $j$ in a graph $G$ is the multiplicative inverse of the shortest path distance between the nodes:
\begin{equation}
    E(G) = \frac{1}{n(n-1)} \sum_{i \neq j \in G} \frac{1}{d_{i,j}}.
\end{equation}

The average global efficiency of a graph is the average efficiency of all pairs of nodes.

\section{Results}
\label{sec:3}

\begin{figure*}[!t]
\centering	
\includegraphics[width=\textwidth]{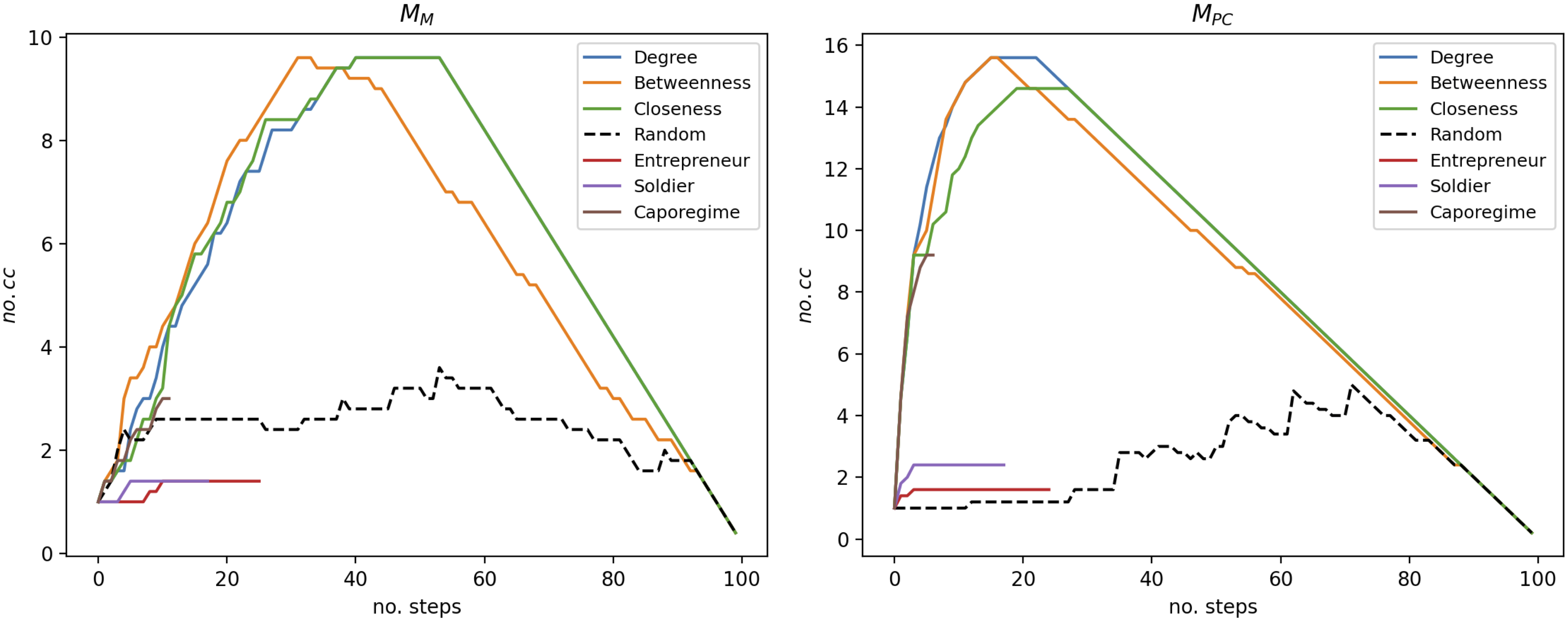}
\caption{Number of connected components in Montagna Meetings and Montagna Phone Calls networks.\label{fig1}}
\end{figure*}

\begin{figure*}[!t]
\centering
\includegraphics[width=\textwidth]{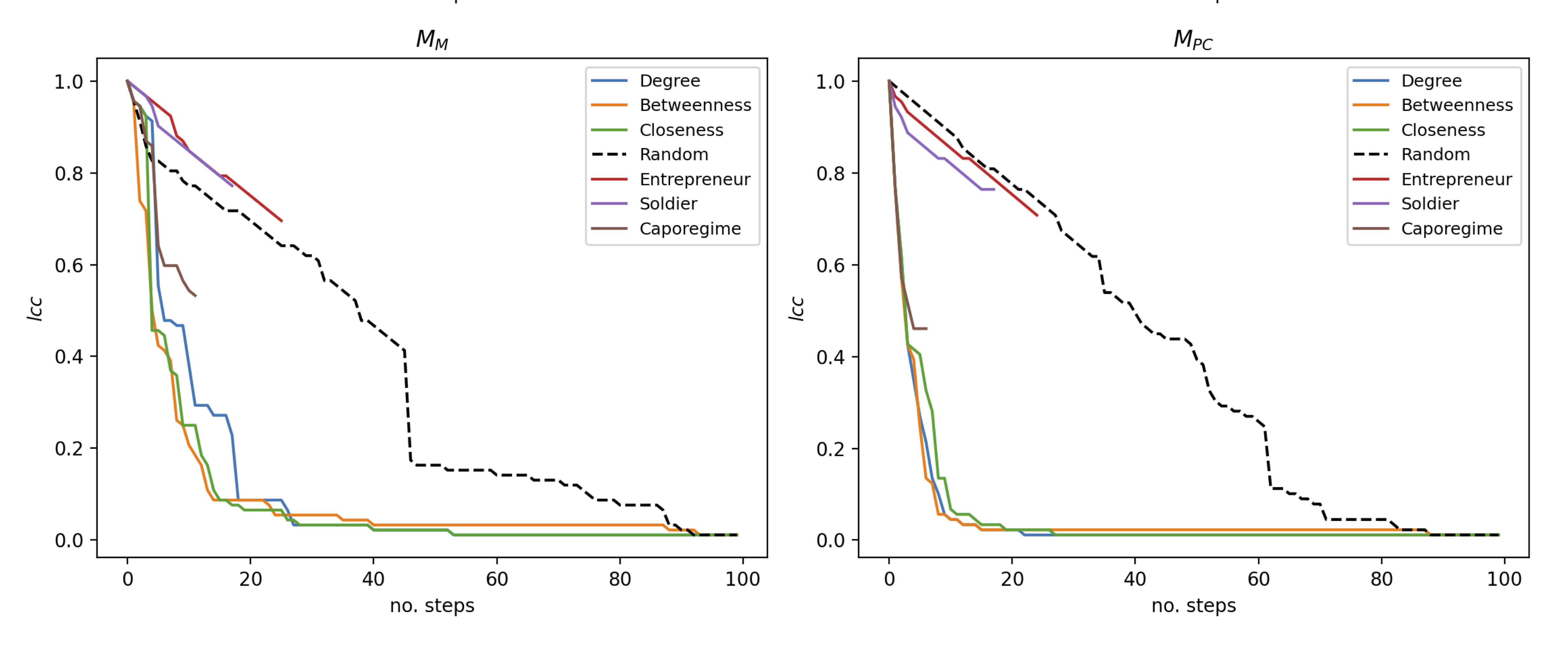}
\caption{Largest connected component in Montagna Meetings and Montagna Phone Calls networks.\label{fig2}}
\end{figure*}

\begin{figure*}[!t]
\centering
\includegraphics[width=\textwidth]{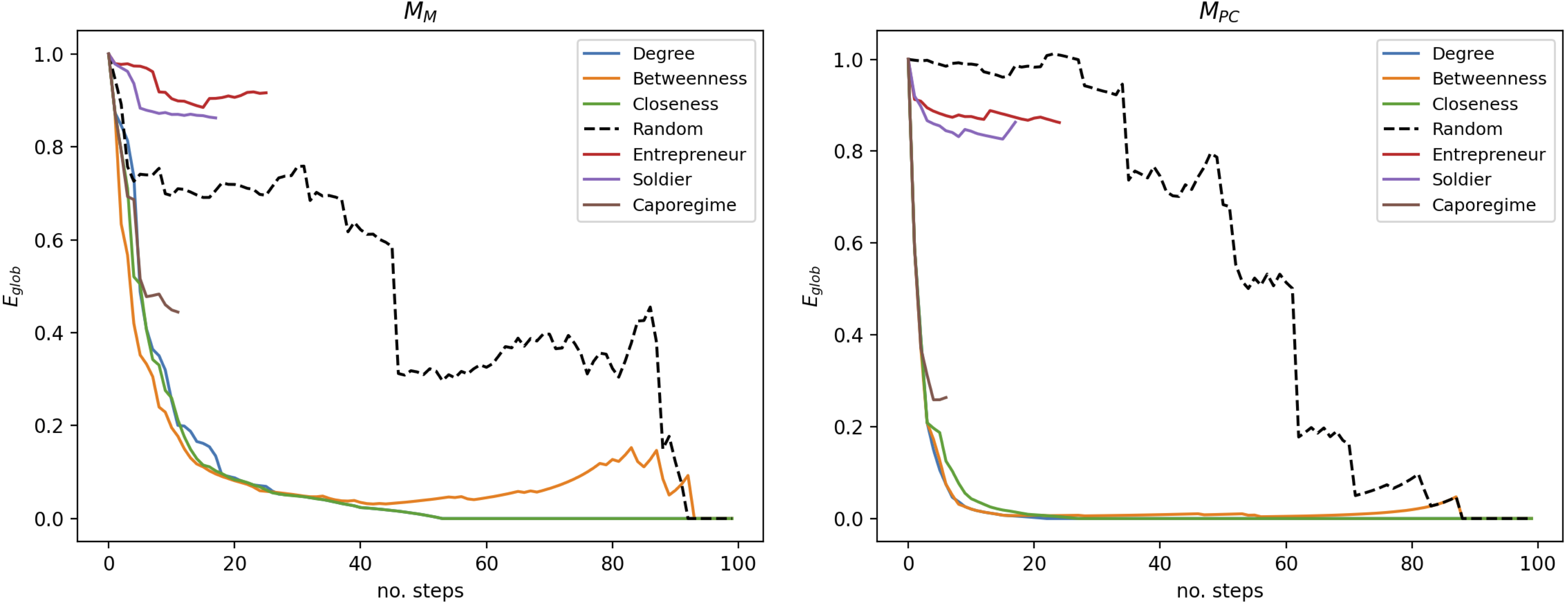}
\caption{Global efficiency in Montagna Meetings and Montagna Phone Calls networks.\label{fig3}}
\end{figure*}

To facilitate comparisons across the disruption strategies described in Subsect.~\ref{sec:distrat}, we plotted three outcome measures on three separate figures: number of connected components (see Figure~\ref{fig1}), largest connected component size (see Figure~\ref{fig2}), and average global efficiency (see Figure~\ref{fig3}). For each plot, the x-axis shows the number of steps performed. At each step, one actor is removed according to the social, human or random approach. 

Our results show that the social capital approach is able to increase the number of connected components, to decrease the size of the largest connected components and the network efficiency in both the Meetings and Phone Calls networks on average by step $20$. 

Random disruption strategy is the least effective. 

The human capital approach is as ineffective as the random one. Unexpectedly, targeting based on entrepreneurs seems to be able to disrupt the networks despite they should have a key role in the Montagna operation. Targeting based on caporegimes represents an exception because it seems to be able of disrupting the networks as the degree, betweenness and closeness targeting.

\begin{figure*}[!t]
\centering	
\includegraphics[width=\textwidth]{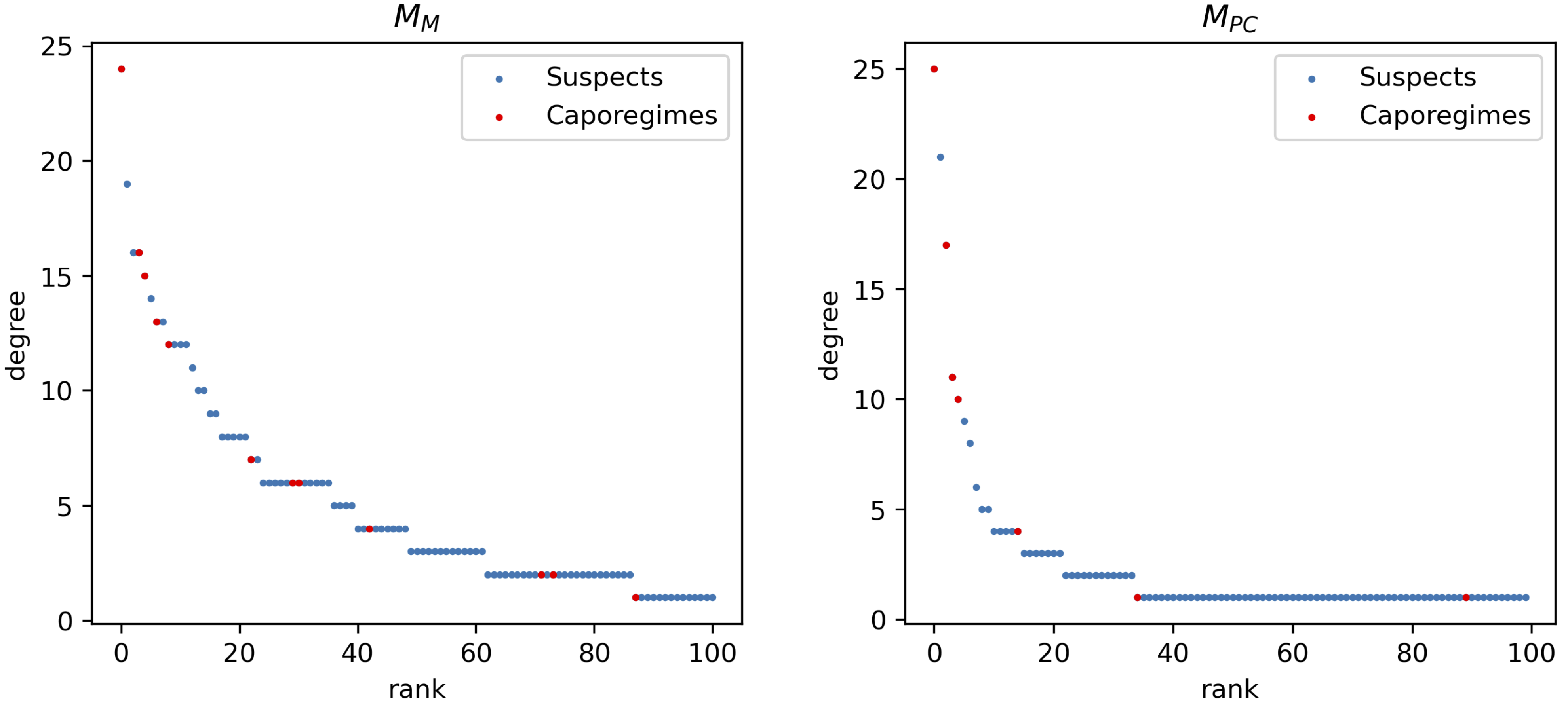}
\caption{Ranking nodes in Montagna Meetings and Montagna Phone Calls networks according to their degree of connectivity.\label{fig4}}
\end{figure*}

Based on this good result about the removal of caporegimes, we decided to rank nodes according to their degree of connectivity, highlighting in red the caporegimes (see Figure~\ref{fig4}). 

Then, we did a different kind of analysis to know:
\begin{enumerate*}[label={(\arabic*)}]
    \item if it is possible to identify a role of a Mafia family as the caporegime on a network model based on the ranking of nodes;
    \item if the application of the random, social and human capital disruption strategies is effective on a network model.
\end{enumerate*}

In one of our previous works~\cite{Ficara2021Green}, we used some popular network models like random networks (\textit{i.e.} the ER~\cite{Erdos1959} model), small-world networks (\textit{i.e.} the WS~\cite{Watts1998} model), and different configurations of scale-free networks (\textit{i.e.} the BA~\cite{Barabasi1999} model) to replicate the topology of our Meetings network. 

Since our experiments identified the BA model as the one which better represents the  criminal networks under study in the present work, we ranked nodes according to their degree of connectivity in two kinds of BA models. We highlighted in red the nodes with the same rank of the caporegimes in the Meetings and Phone Calls networks (\textit{i.e.} the supposed caporegimes). A BA graph of $n$ nodes is grown by attaching new nodes each with $m$ edges that are preferentially attached to existing nodes with high degree. In this study we chose $n=100$ and $m=2$ and $m=3$.

\begin{figure*}[!t]
\centering	
\includegraphics[width=\textwidth]{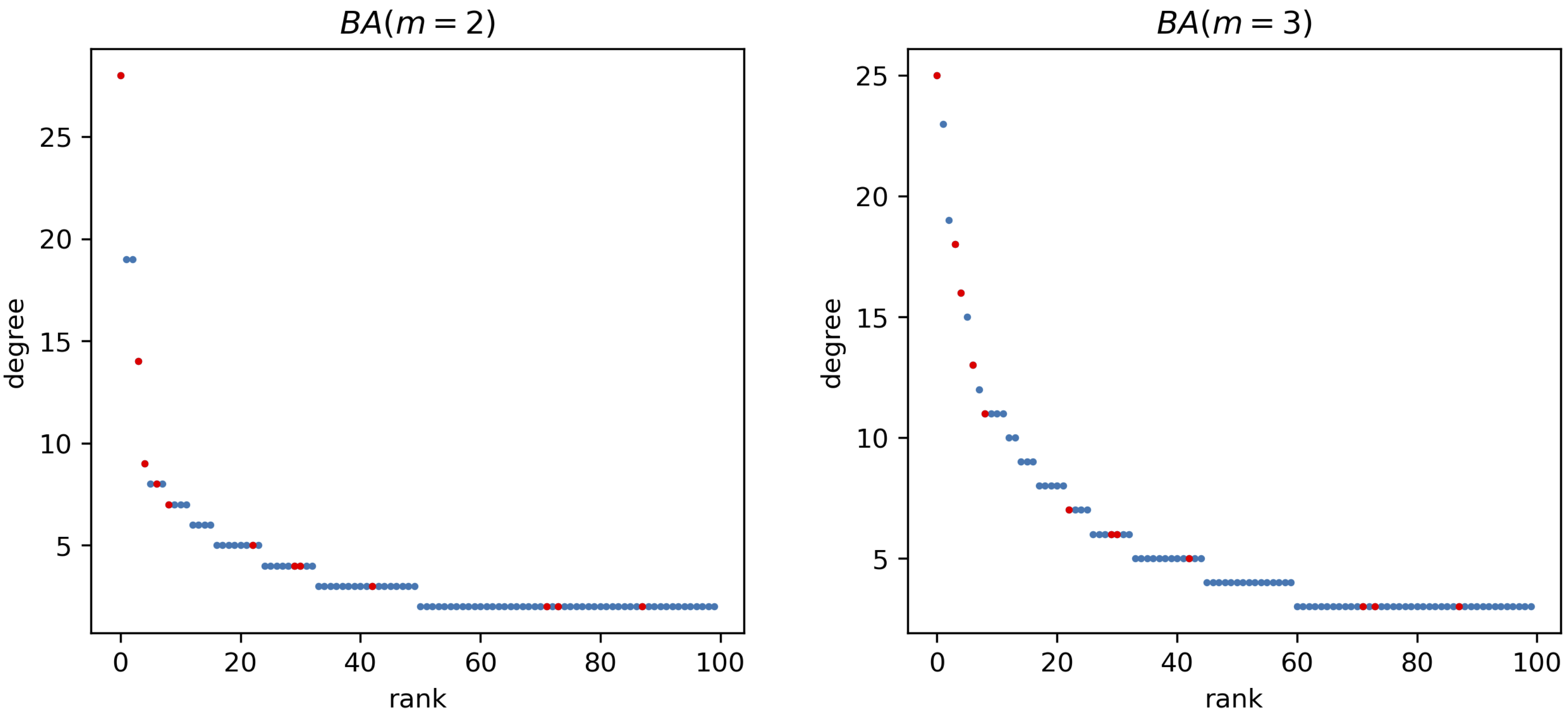}
\caption{Ranking nodes in Barab\'{a}si-Albert models according to their degree of connectivity.\label{fig5}}
\end{figure*}

\begin{figure*}[!t]
\centering	
\includegraphics[width=\textwidth]{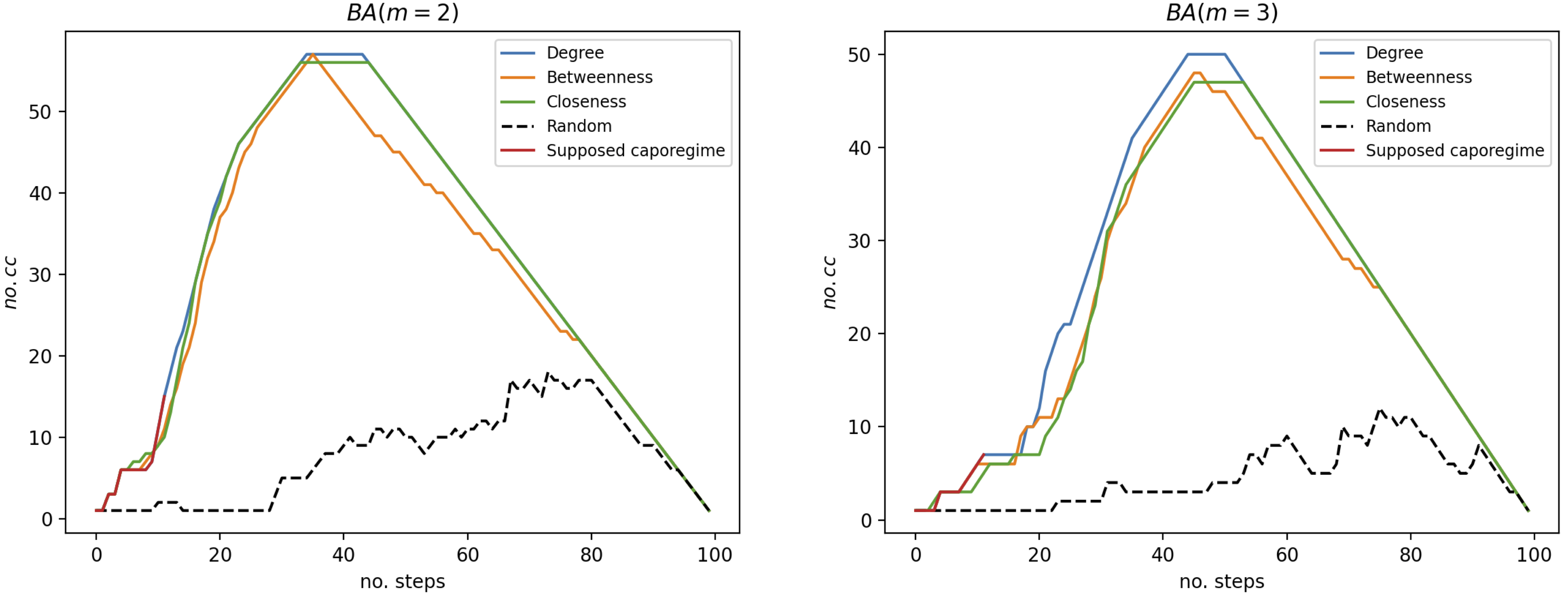}
\caption{Number of connected components in Barab\'{a}si-Albert models.\label{fig6}}
\end{figure*}

\begin{figure*}[!t]
\centering	
\includegraphics[width=\textwidth]{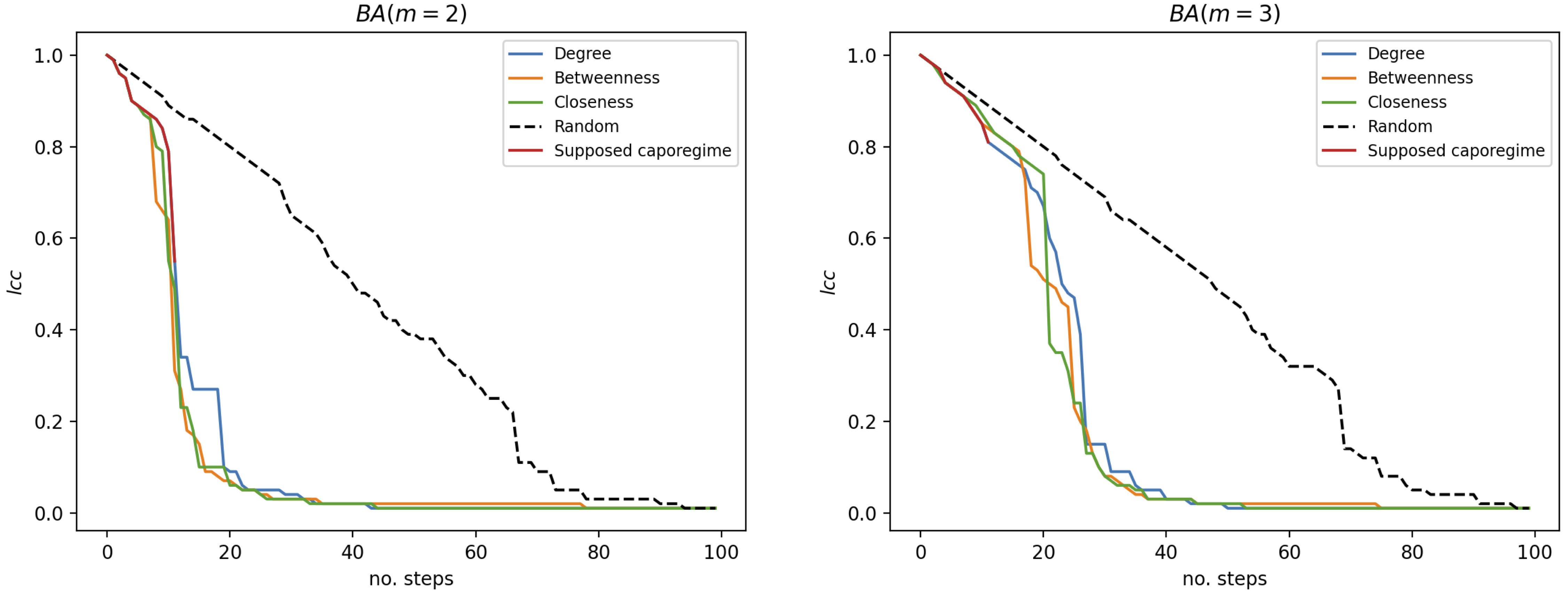}
\caption{Largest connected component in Barab\'{a}si-Albert models.\label{fig7}}
\end{figure*}

\begin{figure*}[!t]
\centering	
\includegraphics[width=\textwidth]{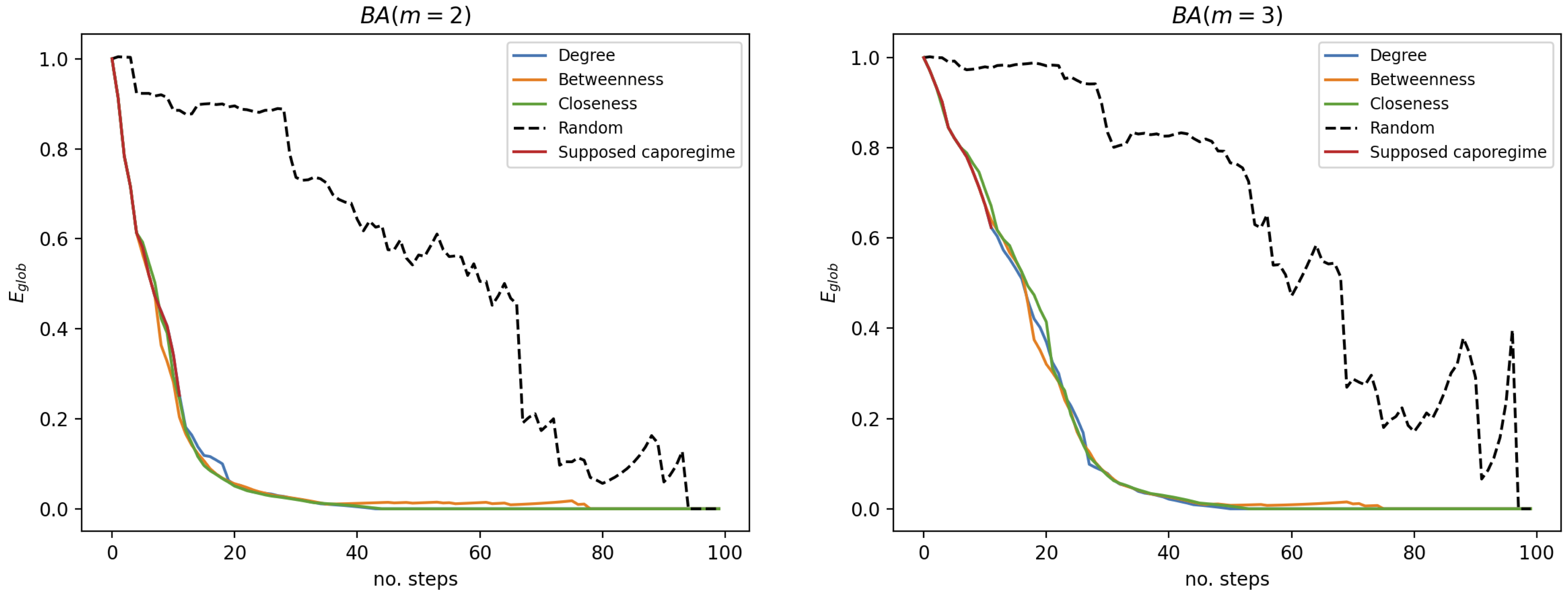}
\caption{Global efficiency in Barab\'{a}si-Albert models.\label{fig8}}
\end{figure*}

Then, we applied our disruption strategies to the BA models. We plotted the three outcome measures on three separate figures: number of connected components (see Figure~\ref{fig6}), largest connected component size (see Figure~\ref{fig7}), and average global efficiency (see Figure~\ref{fig8}). Our results show once again the efficiency of the social capital approach respect to the random one and how the targeting of the supposed caporegimes appears effective as the social capital approach. The efficacy of the removal of the supposed caporegimes also proves that the caporegimes were correctly identified in the network model.
Moreover, the results obtained for the BA graph with $m=2$ are more similar to the one obtained for our criminal networks. The network in fact starts to be dismantled on average by step $20$. The BA model with $m=3$ starts to be dismantled on average by step $30$.

\section{Conclusions}

Application of SNA in criminology has already been applied in the past, as in the study of Mafia networks that have been showed to stand out from other types of criminal networks due to their structure. This study allowed to simulate different types of interventions to disrupt two real criminal Mafia networks. 
Such a framework allowed to test various hypothetical disruption scenarios by comparing three law enforcement interventions that targeted social capital (degree centrality, betweenness centrality, closeness centrality) and three law enforcement strategies that targeted human capital. In case of Mafia networks, such human capital-based strategies target actors belonging to a specific role rather than actors owning specific resources and skills like in other types of criminal networks. In particular, in this study, the human capital-based strategies target respectively the roles of entrepreneur, soldier and caporegime.

A seventh strategy based on random removal was used as a baseline against which to compare the performance of the other six. Such strategy is comparable with opportunistic law enforcement interventions. All strategies based on social capital and human capital were far more effective at disrupting the Mafia network compared with the random one.

Overall, the most effective disruption strategies showed to be the ones that target actors with the highest social capital, with the betweenness centrality-based one to be the best performing among the three. Human capital-based strategies showed to be quite ineffective, with the one targeting caporegimes to perform best among the three. For this reason, another analysis was carried to understand if it would be possible to identify a role in a Mafia networks based on the ranking of the nodes.
Such new analysis was carried to repeat and confirm the experiments on a Barab\'{a}si-Albert model which was shown, from our previous studies, to be the artificial model that best reproduces Mafia networks.
Results showed once again the effectiveness of the social capital-based approaches and the caporegime-based one with respect to the random disruption strategy.

In the real world, the arrest of actors inside a criminal group could cause perturbations inside the network that lead to ties broken and other new created and overall readjustments. The feature of being dynamic and adaptable is a crucial concept tied to criminal networks. In literature, this phenomenon is generally referred to as resilience. Resilience is indeed the ability of such networks to face pressures from LEAs and to reorganize after specific attacks, based on the original topology. For instance, the arrest of actors may threaten trust inside the network, since it may raise suspicions of informants close to the arrested subject, that the arrested individual may have confessed the crime and expose others or that LEAs have other network members under surveillance. Such dynamics could cause a reorganization inside the network.

SNA tools allow to study criminal network resilience. From previous studies it has emerged that after going against strong perturbations, criminal networks succeed in reestablishing their structure, even by substituting the missing members. Future simulations could then model such responses, taking network adaption into account.

One limitation in SNA application in criminology is finally related to the data source and to the modeling used for the simulations. Although criminal data are a common source, such data are vulnerable to error for several reasons. It generally suffers from incompleteness, given the covert nature of criminal networks; incorrectness, due to human errors or deceptions; inconsistency, caused by misleading information. Moreover, gathering complete network data is an impossible task, due to the feature of a criminal network to be dynamic and due to the impossibility to establish its boundaries that are often prone to ambiguity. Finally, there is not a standard method in SNA to turn data into graph, that is a labor-intensive and time-consuming procedure left to the analyst themself.

\bibliographystyle{IEEEtran}
\bibliography{mybibfile}
%


\section{Biography Section}

\begin{IEEEbiographynophoto}{Annamaria Ficara} is a Teaching Assistant at the University of Messina, Italy. She has a PhD in Mathematics and Computational Sciences from the University of Palermo, Italy. Her main research interests are in network science, social network analysis, criminal networks and complex systems. Contact her at \texttt{aficara@unime.it}.
\end{IEEEbiographynophoto}

\begin{IEEEbiographynophoto}{Francesco Curreri} is a PhD Student in Mathematics and Computational Sciences at the University of Palermo, Italy. His main research interests are in system identification and nonlinear systems modeling, soft sensors, neural models, social network analysis and criminal networks. Contact him at \texttt{fcurreri@unime.it}.
\end{IEEEbiographynophoto}

\begin{IEEEbiographynophoto}{Giacomo Fiumara} is an Associate Professor of Computer Science at the University of Messina, Italy since 2009. In 1993 he took his PhD in Physics. His research interests include network science, criminal networks, simulations of model systems. He has published more than 80 papers in international journals and conference proceedings. He is member of various conference PCs. Contact him at
\texttt{gfiumara@unime.it}.
\end{IEEEbiographynophoto}

\begin{IEEEbiographynophoto}{Pasquale De Meo} is an Associate Professor of Computer Science at the Department of Ancient and Modern Civilizations at the University of Messina, Italy. His main research interests are in the area of social networks, recommender systems, and user profiling. De Meo has a PhD in Systems Engineering and Computer Science from the University of Calabria. He has been the Marie Curie Fellow at Vrije Universiteit Amsterdam. Contact him at \texttt{pdemeo@unime.it}.
\end{IEEEbiographynophoto}

\vfill

\end{document}